\documentclass[pra,reprint,bibnotes,floatfix]{revtex4-2}  

\usepackage[T1]{fontenc}
\usepackage[utf8]{inputenc} 
\usepackage[australian]{babel}
\usepackage[a4paper,centering,hmargin=1.75cm,vmargin=2cm]{geometry} 
\usepackage{amsmath,amssymb,graphicx,bm,microtype} 
\usepackage[dvipsnames]{xcolor}
\usepackage[colorlinks,allcolors=blue!50!black]{hyperref}
\usepackage[all]{hypcap} 
\usepackage{braket,cleveref,enumitem,titlesec,siunitx,newfloat}
\titleformat{\paragraph}[runin]{\bfseries}{}{0pt}{}[.]
\DeclareFloatingEnvironment[
  fileext = loa,
  listname = {List of algorithms},
  name = Algorithm
]{algorithm}
\newcommand{\abs}[1]{\lvert #1 \rvert}
\renewcommand{\vec}[1]{\bm{\mathrm{#1}}}
\let\Re\relax \DeclareMathOperator{\Re}{Re}

\begin{document}

\title{Even a little delocalisation produces large kinetic enhancements of charge-separation efficiency in organic photovoltaics}

\author{Daniel Balzer}
\affiliation{School of Chemistry and University of Sydney Nano Institute, University of Sydney, NSW 2006, Australia}

\author{Ivan Kassal}
\email[Email: ]{ivan.kassal@sydney.edu.au}
\affiliation{School of Chemistry and University of Sydney Nano Institute, University of Sydney, NSW 2006, Australia}

\begin{abstract}
In organic photovoltaics, charges can separate efficiently even if their Coulomb attraction is an order of magnitude greater than the available thermal energy. Delocalisation has been suggested to explain this fact, because it could increase the initial separation of charges in the charge-transfer~(CT) state, reducing their attraction. However, understanding the mechanism requires a kinetic model of delocalised charge separation, which has proven difficult because it involves tracking the correlated quantum-mechanical motion of the electron and the hole in large simulation boxes required for disordered materials. Here, we report the first three-dimensional simulations of charge-separation dynamics in the presence of disorder, delocalisation, and polaron formation, finding that even slight delocalisation, across less than two molecules, can significantly enhance the charge-separation efficiency, even starting with thermalised CT states. Importantly, delocalisation does \emph{not} enhance efficiency by reducing the Coulomb attraction; instead, the enhancement is a kinetic effect produced by the increased overlap of electronic states. 
\end{abstract}

\maketitle

\section{Introduction}

\begin{figure}
    \centering
    \includegraphics[width=\columnwidth]{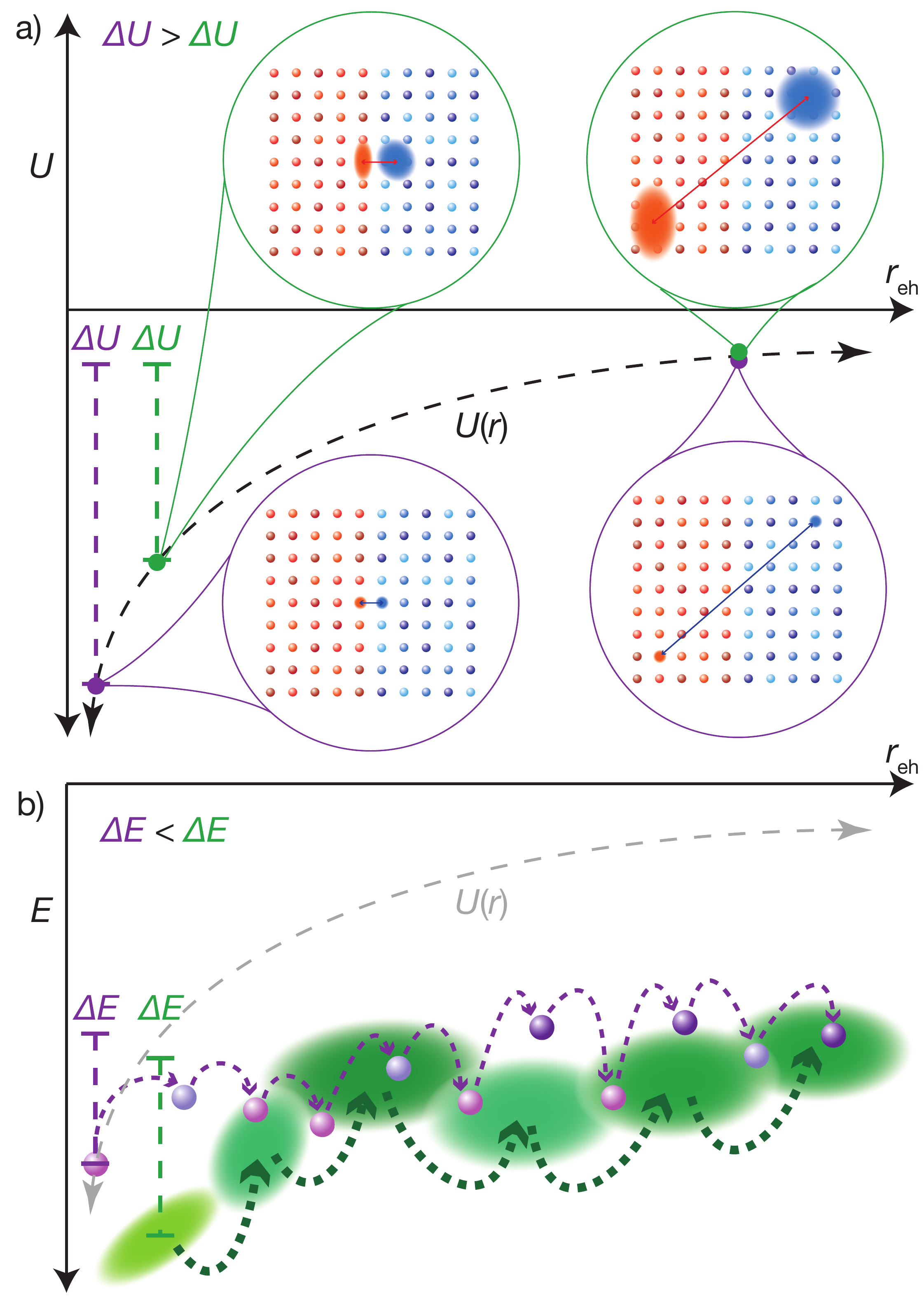}
    \caption{\textbf{Mechanism of delocalisation-enhanced  charge separation.}
    \textbf{a)} Charge separation is typically modelled (purple insets) by assuming that the charges are localised onto individual molecules or sites (spheres) by disorder (different shades). When both charges are localised at the donor-acceptor interface, they have a small electron-hole separation $r_\mathrm{eh}$ and a large Coulomb attraction $U(r)$. It has been proposed that delocalisation of charges across many molecules (green insets) facilitates their separation by increasing their initial separation and decreasing their Coulomb attraction. 
    \textbf{b)} While delocalisation does decrease the Coulomb binding energy ($\Delta U$), it increases the overall binding energy ($\Delta E$), meaning that a reduction in the CT-state binding energy is not the cause of delocalisation enhancements. Instead, delocalisation increases the overlaps between electronic states, allowing delocalised charges (green clouds) to hop further and faster than localised ones (purple dots).
    } 
    \label{fig:delocalised_charge_separation}
\end{figure}

The precise mechanism of how charges in many organic photovoltaics (OPVs) separate with near-perfect efficiency remains unclear, especially considering that their Coulomb attraction can be more than an order of magnitude larger than the available thermal energy~\cite{Park2009,Few2015}. OPVs typically contain a heterojunction of electron-donor and electron-acceptor materials. Upon excitation, an exciton is created in either material and diffuses to the donor-acceptor interface, where charge transfer can form an interfacial charge-transfer (CT) state (left of \cref{fig:delocalised_charge_separation}a). For the photogeneration process to continue, the charges must escape their Coulomb attraction and separate (right of \cref{fig:delocalised_charge_separation}a). In organic semiconductors with the typical dielectric constants of 3--4, charges separated by \SI{1}{nm} across the interface experience a Coulomb attraction of 360--\SI{480}{meV}, more than an order of magnitude more than the thermal energy $k_\mathrm{B}T=\SI{25}{meV}$. 

The efficient separation is commonly attributed to charge delocalisation, an argument supported by the experimental observation of efficient and fast separation via delocalised CT states~\cite{Bakulin2012,Jailaubekov2013,Grancini2013,Gelinas2014,Falke2014,Tamai2017}.
On this view, the delocalisation increases the initial separation of charges in the CT state, reducing the Coulomb attraction and the energetic barrier that needs to be overcome (\cref{fig:delocalised_charge_separation}a). However, the concept of a kinetic barrier to charge separation is problematic, because separation is not a one-step process from a CT state to a separated state, but a multiple-step process through many states, where each transition has its own energetic barrier (\cref{fig:delocalised_charge_separation}b). Furthermore, the barrier can completely disappear in free energy when entropy and disorder are considered~\cite{Hood2016}, and actually increase when delocalisation is included~\cite{Gluchowski2018}. Therefore, any charge-separation efficiency improvement caused by delocalisation must come from non-equilibrium kinetic effects, not purely energetic considerations~\cite{Gluchowski2018,Shi2017}.

Many kinetic models of charge separation have included charge delocalisation, with most of them finding an increased efficiency when delocalisation is considered~\cite{Sun2014,Smith2014,Smith2015,Abramavicius2016,DAvino2016,Jankovic2017,Jankovic2018,Yan2018,Kato2018,Caruso2012,LeeTroisi2015,Kelly2020, Deibel2009,Nenashev2011,Schwarz2013,Tscheuschner2015,Athanasopoulos2017,Athanasopoulos2019}. These approaches have ranged from quantum-mechanical descriptions of delocalisation in disordered materials~\cite{Sun2014,Smith2014,Smith2015,Abramavicius2016,DAvino2016,Jankovic2017,Jankovic2018,Yan2018} to phenomenological treatments that include delocalisation in an effective way~\cite{Deibel2009,Nenashev2011,Schwarz2013,Tscheuschner2015,Athanasopoulos2017,Athanasopoulos2019}.
However, different approaches have used approximations that limit their range of applicability in three important ways.
First, approaches that do not include static disorder are applicable to organic crystals, but not disordered organic semiconductors. Second, approaches that treat delocalisation in an effective way tend to not include the complete quantum-mechanical description required for accurately predicting the effects and extent of delocalisation. Finally, approaches that do not adequately describe the coupling of the charges to their environment can fail to describe the formation of polarons, which can localise the states significantly~\cite{Rice2018}, meaning that those approaches can overestimate delocalisation and efficiency enhancements.  
Therefore, an accurate kinetic model of charge separation in OPVs must include these three important ingredients: disorder, a quantum-mechanical treatment of delocalisation, and polaron formation.

The most complete kinetic models of charge separation have included all three important ingredients~\cite{Tamura2013,Bittner2014,Jankovic2020}. However, they remain computationally expensive, limiting their application to small systems, low dimensions, or short times. Tamura and Burghart used an atomistic approach, solving the time-dependent Hamiltonian using multiconfigurational time-dependent Hartree (MCTDH)~\cite{Tamura2013,Huix-Rotllant2015,Polkehn2018} to predict the ultrafast separation of an interfacial exciton over about 100 fs, attributing the efficient separation to vibronically hot CT states and to charge delocalisation reducing the Coulomb attraction. While atomistic approaches do not require adjustable parameters, they are expensive and therefore limited to relatively few states and short times. Bittner and Silva~\cite{Bittner2014} and Jankovi\'c and Vukmirovi\'c~\cite{Jankovic2020} both parametrised model Hamiltonians, which allows for longer simulations on more states, as fewer degrees of freedom are tracked. While their approaches are computationally cheaper, they are still limited to two- and one-dimensional simulations, respectively. Both approaches include the formation of polarons using the polaron transformation. Bittner and Silva~\cite{Bittner2014} studied the separation of charges in a two-dimensional heterojunction using a time-dependent Schr\"odinger equation. They observe fast separation into free polarons, due to charges tunneling from the initial exciton in under $\SI{35}{fs}$ (rather than proceeding through CT states), due to resonance between excitons and free polarons. However, the environment was treated as a bath of extended, shared phonons (which is common for crystalline systems), rather than as local molecular vibrations, which is more appropriate for disordered molecular systems~\cite{MayKuhn}. Jankovi\'c and Vukmirovi\'c~\cite{Jankovic2020} modelled charge separation in a one-dimensional system over long times to study the competition of delocalisation, disorder and polaron effects. They found that separation proceeds slowly through thermalised CT states; after excitons form CT states in 1--\SI{10}{ps}, the latter separate in about \SI{1}{ns}. However, rates were calculated using modified Redfield theory, which requires small off-diagonal system-bath couplings, an assumption that is not met in many materials, including some organic semiconductors, where strong system-bath coupling alone can localise the electronic states~\cite{Jesenko2014}. 

The best kinetic models have remained computationally expensive for two main reasons. First, correctly describing delocalisation requires a quantum-mechanical treatment, which is difficult in disordered materials, where periodic boundary conditions must be replaced with large simulation boxes. Second, charge separation is a two-body problem involving the correlated motion of an electron and a hole, meaning that the Hilbert space size is roughly the square of a single-body calculation. As a result, a quantum-mechanical treatment has so far proved intractable in three dimensions~\cite{Few2015}. In order to study charge separation over long times, in large systems, and in three dimensions, we require a more efficient approach that still meets all requirements.

A logical starting point for understanding delocalisation in charge separation is extending kinetic models that describe the transport of partially delocalised single particles~\cite{Oberhofer2017,Giannini2018,Giannini2019,Giannini2020,Ellis2021,Varvelo2021,Lee2015,Balzer2020,vukmirovic2009,mladenovic2015,Campaioli2021}. However, even this task is difficult when disorder, delocalisation, and polaron formation are required~\cite{Oberhofer2017}.
The best recent approaches to tackle the single-particle problem are all effective-Hamiltonian methods, and include fragment orbital-based surface hopping (FOB-SH)~\cite{Giannini2018,Giannini2019,Giannini2020,Ellis2021}, where the Hamiltonians are parametrised using an atomistic approach requiring no adjustable parameters, as well as adHOPS~\cite{Varvelo2021}, which emphasises the non-perturbative treatment of the system-environment couplings using hierarchy equations of motion. The computational cost of these approaches has limited their application to short times in two dimensional systems or large one-dimensional systems.

Recently, we introduced delocalised kinetic Monte Carlo (dKMC), the first three-dimensional model of charge and exciton transport that includes disorder, delocalisation, and polaron formation~\cite{Balzer2020}. dKMC is a computationally improved version of secular polaron-transformed Redfield theory (sPTRE)~\cite{Lee2015}, which is the stationary and secular limit of MC-CRET~\cite{Jang2008,Jang2009,Jang2011}, and one example of a second-order polaronic master equation~\cite{Nazir2009,McCutcheon2011,Kolli2011,McCutcheon2011_2,McCutcheon2011_3,Pollock2013,Xu2016}. Because sPTRE is in the polaron frame, which changes as a function of system-bath coupling, it can describe polaron transport in the intermediate regime, while also reproducing the well-known hopping and band conduction extremes~\cite{Lee2015}. dKMC shows that transport in moderately disordered materials is that of charges hopping between partially delocalised electronic states and that even a small amount of delocalisation can increase carrier mobilities dramatically~\cite{Balzer2020}.

\begin{figure*}
    \centering
    \includegraphics[width=0.95\textwidth]{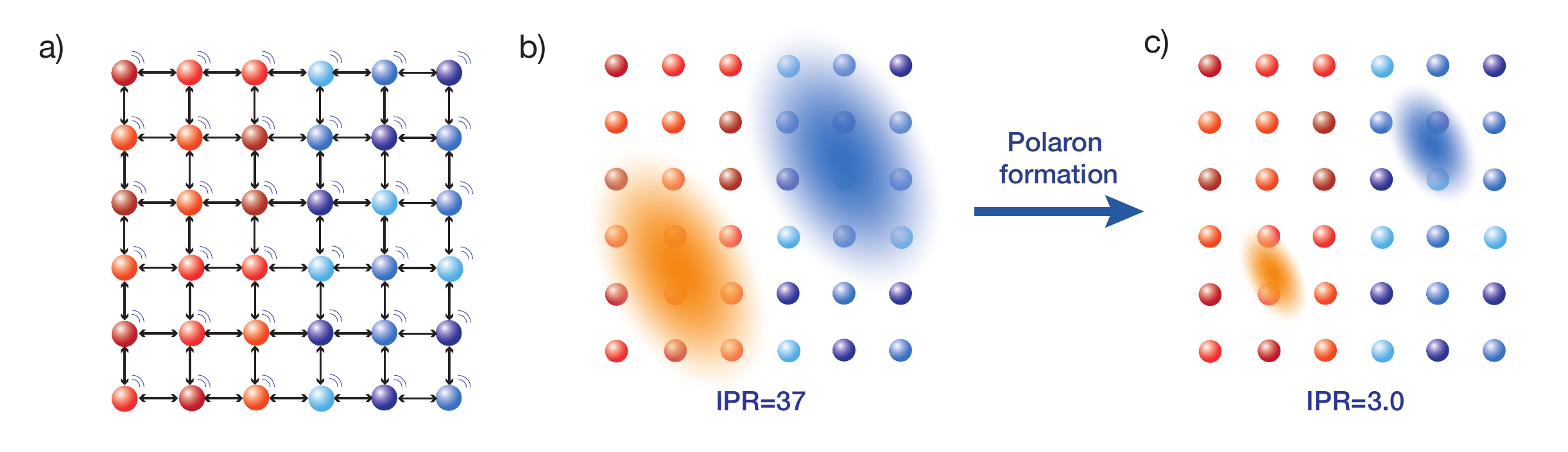}
    \caption{\textbf{Components of the dKMC model.} \textbf{a)} A heterojunction is modelled as a lattice of sites with disordered energies (different shades), representing acceptor (orange) and donor (blue) molecules. Each site is coupled to an environment (motion lines) and to its neighbours. 
    \textbf{b)} Delocalisation of the electronic states is found by diagonalising the system's Hamiltonian, with the eigenstates representing the simultaneous position of the electron (in the acceptor, blue) and the hole (in the donor, orange). 
    \textbf{c)} Polaron formation further localises the states.
    Inverse participation ratios (IPRs) are shown for $J=\SI{30}{meV}$ and $\sigma=\SI{150}{meV}$.
    }
    \label{fig:model}
\end{figure*}

Here, we extend dKMC to make the charge-separation problem computationally accessible, allowing the first three-dimensional simulation of the dynamics and efficiency of charge separation in the presence of disorder, delocalisation, and polaron formation. We use it to show that small amounts of delocalisation can produce large efficiency enhancements, even for thermalised CT states. Contrary to the common hypothesis, these delocalisation enhancements are not a consequence of a reduction in the initial Coulomb binding energy. Rather, delocalisation actually increases the total binding energy, and the efficiency enhancements are a kinetic effect caused by greater overlaps between electronic states, which allow charges to move further and faster (\cref{fig:delocalised_charge_separation}b). All of the approximations used in this work are conservative, chosen to underestimate the extent and role of delocalisation; in the end, we discuss ways to relax the approximations, which could lead to an even greater role for delocalisation.

\section{Results\label{sec:results}}

\paragraph{Model}
Our approach is based on delocalised Kinetic Monte Carlo (dKMC)~\cite{Balzer2020}. Here, we summarise the changes used to extend dKMC to the two-body charge-separation problem, which are detailed in \cref{sec:theory}.

The charge-separation problem is described by the Hamiltonian
\begin{equation}
    \label{eq:H}
    H = H_\mathrm{S} + H_\mathrm{B} + H_\mathrm{SB},
\end{equation}
whose components describe the system $H_\mathrm{S}$, the bath $H_\mathrm{B}$, and the interaction between them $H_\mathrm{SB}$.

We model the system using a tight-binding model of a $d$-dimensional cubic lattice containing $N^d/2$ donor sites next to $N^d/2$ acceptor sites, each representing a molecule or part of a molecule (\cref{fig:model}a). Each site is assigned a HOMO and LUMO energy, so that an electron occupying that site will have the LUMO energy and a hole the HOMO energy. These energies are assumed to be disordered in order to model the different local environments around different molecules, which arise from static variations in the spacing and orientation of molecules. The HOMO and LUMO energies are drawn randomly from Gaussian distributions $\mathcal{N}(E_0^\mathrm{HOMO},\sigma)$ and $\mathcal{N}(E_0^\mathrm{LUMO},\sigma)$, where the energetic disorder $\sigma$ is assumed to be equal for the HOMOs and the LUMOs~\cite{Bassler1993}. 

In the two-body problem, the Hilbert space consists of ordered pairs of sites, where $\ket{m,n}$ represents an electron on donor site $m$ and the hole on acceptor site $n$. The energy of this pair is $E_{m,n}=E_m^\mathrm{LUMO}-E_n^\mathrm{HOMO}+U(r)$, where $E_m^\mathrm{LUMO}$ is the LUMO energy of site $m$, $E_n^\mathrm{HOMO}$ is the HOMO energy of site $n$ and $U(r)$ is the Coulomb potential of charges separated by a distance $r$,
\begin{equation}
    U(r)=-\frac{e^2}{4\pi\varepsilon_0 \varepsilon_r r},
\end{equation}
where $e$ is the elementary charge, $\varepsilon_0$ is the vacuum permittivity and $\varepsilon_r$ is the dielectric constant (here, taken to be $\varepsilon_r=3.5$). 

A pair of sites is electronically coupled to other pairs of sites that can be obtained by moving either the electron or the hole, but not both, giving the system Hamiltonian 
\begin{multline}
    H_\mathrm{S} = \sum_{m\in D,n \in A} E_{mn} \ket{m,n}\bra{m,n} \\ + \sum_{m\neq m' \in D,n \in A} J_{mn,m'n} \ket{m,n}\bra{m',n}\\ + \sum_{m \in D, n\neq n' \in A} J_{mn,mn'} \ket{m,n}\bra{m,n'}.
    \label{eq:H_S}
\end{multline}
In general, the couplings $J_{mn,m'n}$ and $J_{mn,mn'}$ can be disordered or long range; however, we assume only constant nearest-neighbour couplings with strength $J$.

\begin{figure*}
    \centering
    \includegraphics[width=\textwidth]{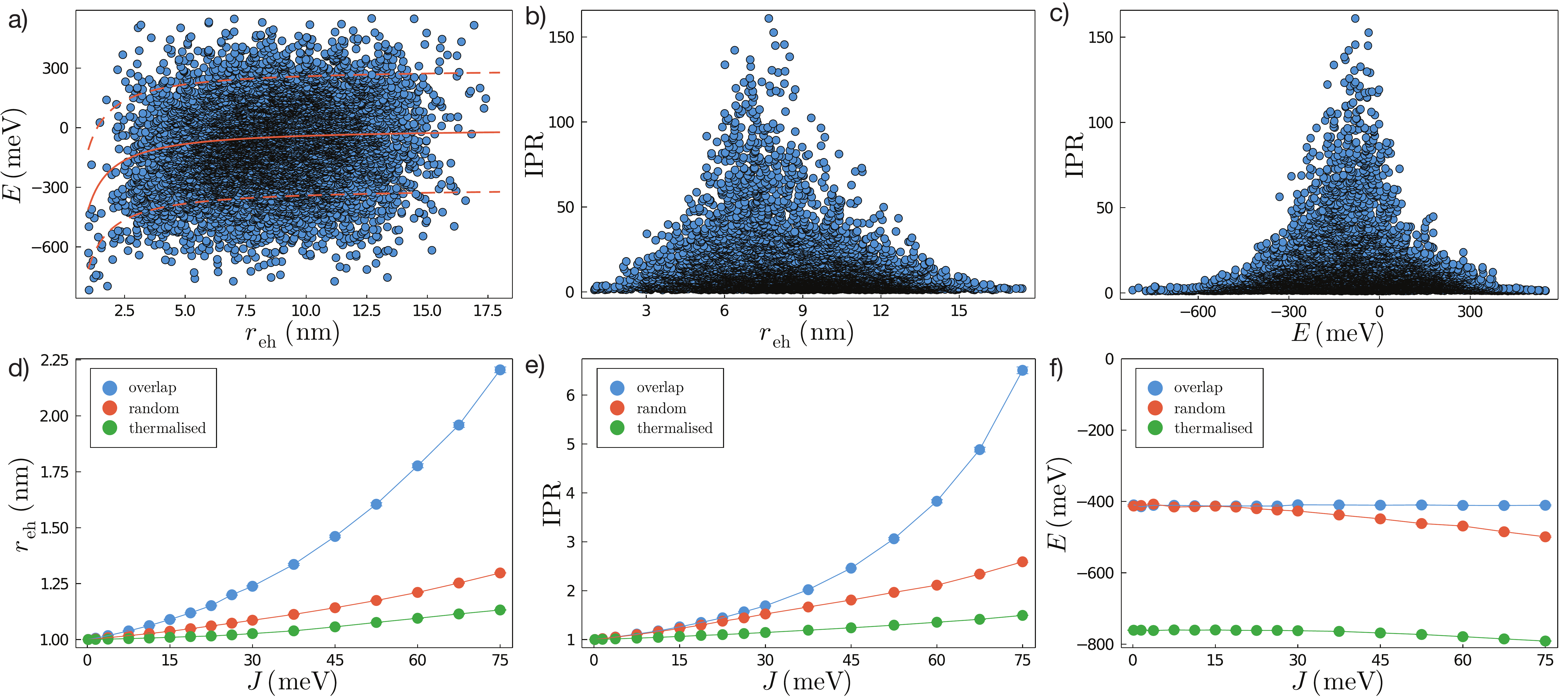}
    \caption{\textbf{Properties of the polaronic states.}
    In panels a--c, the dots represent the polaronic states of a 2D heterojunction with $J=\SI{75}{meV}$ and $\sigma=\SI{150}{meV}$. 
    \textbf{a)} The Coulomb interaction stabilises states with small electron-hole separations (orange line is the Coulomb potential and the dashed lines are $\pm 2\sigma$ on either side). At any $r_\mathrm{eh}$, including for CT states, there are states with a wide range of energies. 
    \textbf{b)} States with small $r_\mathrm{eh}$ (and, therefore, large Coulomb attractions) are more localised. 
    \textbf{c)} Low-energy states are more localised, while the more delocalised states lie closer to the middle of the density of states.
    \textbf{d--f)} The average separation, IPR, and energy of the three kinds of initial CT states, as a function of the electronic coupling $J$. 
    Overlap CT states are chosen based on each state's overlap with interfacial CT site pairs, producing states that are the most separated, delocalised, and highest in energy. 
    Random CT states are chosen uniformly from the states with the smallest $r_\mathrm{eh}$.
    Thermalised CT states are chosen from the states with the smallest $r_\mathrm{eh}$ in proportion to their Boltzmann factor, producing initial states that are the least separated, least delocalised, and lowest in energy.
    The error bars are the standard errors of the mean.
    }
    \label{fig:states}
\end{figure*}

To model the bath, every site is assumed to have an identical, independent bath of harmonic oscillators representing molecular vibrations, which is the usual assumption for disordered molecular materials~\cite{kohler2015textbook,MayKuhn}. Then, the bath Hamiltonian is given by
\begin{equation}
    \label{eq:H_B}
    H_\mathrm{B}=\sum_{m\in D,k}\omega_{mk} b^\dag_{mk}b_{mk} + \sum_{n\in A,k} \omega_{nk} b^\dag_{nk}b_{nk},
\end{equation}
where $\omega_{nk}$ is the frequency of the $k$th bath mode at the $n$th site, and $b^\dag_{nk}$ and $b_{nk}$ are the corresponding creation and annihilation operators. 

The system-bath interaction is described by a coupling of every site to its bath modes. We assume a linear coupling of strength $g_{nk}$ between site $n$ and bath mode $k$, so that the interaction Hamiltonian is
\begin{multline}
    \label{eq:H_SB}
    H_\mathrm {SB} = \sum_{m \in D,k} g_{mk}\ket{m,n}\bra{m,n}(b^\dag_{mk} + b_{mk}) \\ + \sum_{n \in A,k} g_{nk}\ket{m,n}\bra{m,n}(b^\dag_{nk} + b_{nk}).
\end{multline}

The formation of polarons (quasi-particles containing a charge and the distortion it causes to the bath~\cite{Frolich1954,Holstein1959}), is described by applying the polaron transformation to $H$, which displaces the bath modes using the state-dependent displacement operator~\cite{Grover1971}
\begin{multline}
    e^S = \exp{\Bigg(\sum_{m\in D,k} \frac{g_{mk}}{\omega_{mk}}\ket{m}\bra{m}(b^\dag_{mk}-b_{mk})\Bigg)} \\
    \otimes \exp{\Bigg(\sum_{n\in A,k}\frac{g_{nk}}{\omega_{nk}}\ket{n}\bra{n}(b^\dag_{nk}-b_{nk})\Bigg)}.
    \label{eq:polaron}
\end{multline}
Using the polaron transformation has two benefits. First, it reduces the system-bath coupling by absorbing most of the interaction into the polaron itself, allowing the residual interactions to be treated perturbatively~\cite{Grover1971}. Second, it reduces the electronic couplings within the system, giving smaller polarons~\cite{Rice2018} and reducing the complexity of dKMC calculations~\cite{Balzer2020}.

The results of applying the polaron transformation to $H$ are given in \cref{sec:theory}. The polaron-transformed system Hamiltonian $\tilde{H}_\mathrm{S}$ is then diagonalised to find the joint polaron states of the electron and the hole. The extent of the delocalisation of polaron state $\nu$ can be quantified using the inverse participation ratio, 
\begin{equation}
    \label{eq:IPR}
    \mathrm{IPR}_\nu = \Bigg(\sum_{m\in D,n\in A} \abs{\braket{m,n|\nu}}^4\Bigg)^{-1},
\end{equation}
which describes roughly how many pairs of sites $(m,n)$ the state $\nu$ is delocalised over. The IPRs calculated using \cref{eq:IPR} depend on all the parameters of the model; in particular, IPRs increase with electronic coupling $J$ and decrease with disorder $\sigma$ and system-bath coupling $g$. As shown in \cref{fig:states}, the Coulomb interaction stabilises (\cref{fig:states}a) and localises (\cref{fig:states}b--c) states in which the electron and the hole are close together, such as CT states.

We begin dKMC simulations with the charges in a CT state. Because of disorder, there are many CT states of various energies and multiple plausible ways they could be chosen, as reflected in the debate about whether separation proceeds through low-energy, thermalised CT states or high-energy, hot CT states. To examine various possibilities, we choose the initial CT state in one of three ways. 
\textit{Random CT states} are chosen uniformly at random from the $N_\mathrm{CT}$ polaron states with the smallest electron-hole separations, where $N_\mathrm{CT}$ is set equal to the number of interfacial pairs on the lattice. 
\textit{Thermalised CT states} are chosen from the same $N_\mathrm{CT}$ most-closely bound polaron states, but in proportion to their Boltzmann factors.
Finally, \textit{overlap CT states} are chosen from all the polaron states, in proportion to their overlap with interfacial pairs, $\abs{\sum_{(m,n)\in \mathrm{CT}}\bra{m,n}\nu\rangle}^2$. This choice mimics states prepared by charge transfer from an interfacial exciton, which would be most strongly coupled to interfacial pairs. 
Among the three initial conditions, thermalised CT states are, on average, lowest in energy, most bound, and least delocalised, while overlap CT states are at the opposite extreme (\cref{fig:states}d--f).

Starting with any of these initial CT states, dKMC is a form of kinetic Monte Carlo (KMC) that uses polaron-frame Redfield rates, computed in \cref{sec:theory}, to evolve trajectories through the polaron states. As in regular KMC (which instead uses Marcus~\cite{Marcus1956} or Miller-Abrahams~\cite{Miller1960} rates), the next state is chosen with a probability proportional to the rate of transfer to the state~\cite{Coropceanu,Clarke2010,Few2015,kohler2015textbook,Oberhofer2017}. 
The hopping continues until a terminating condition is reached. If the polarons are separated by more than a chosen separation distance $r_\mathrm{sep}$, which we set to $\SI{5}{nm}$, the charge separation is considered successful. By contrast, the charge separation is considered failed if the polarons recombine or if the number of hops exceeds the limit $n_\mathrm{hops}$, which we set to 2000. The latter assumption avoids infinite loops, such as hopping between two low-lying states, where the polarons would probably eventually recombine.
Finally, the internal quantum efficiency (IQE) is calculated as the fraction of trajectories where the polarons separate, averaged over many realisations of disorder.

\paragraph{Delocalisation increases charge-separation efficiency}

\begin{figure}
    \centering
    \includegraphics[width=\columnwidth]{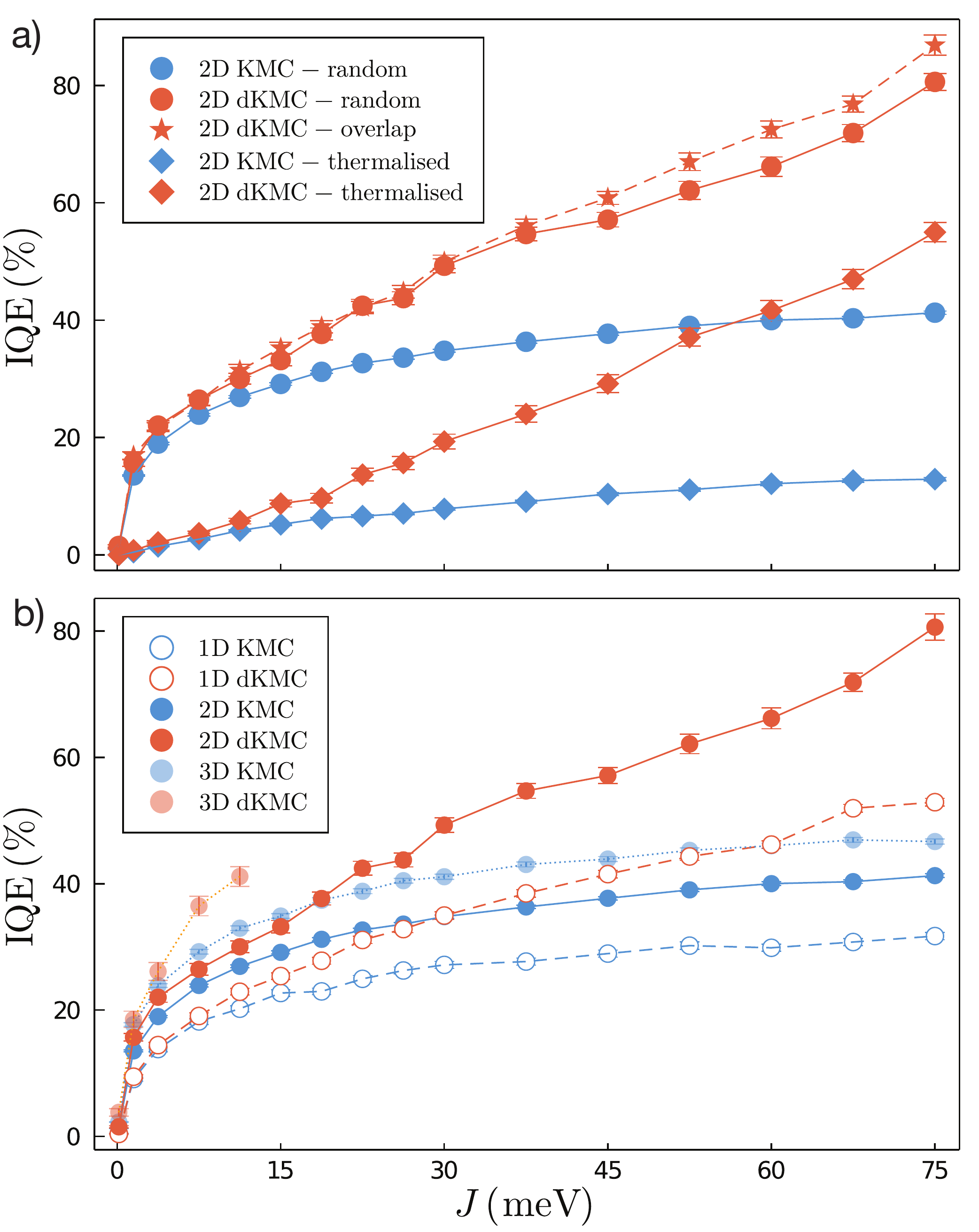}
    \caption{\textbf{Delocalisation and dimensionality increase charge-separation efficiency.}
    \textbf{a)} Internal quantum efficiencies (IQE) of charge separation, for a system with $\sigma=\SI{150}{meV}$ and varying electronic coupling $J$, modelled using regular KMC (blue) and dKMC (orange). When the states are localised (small $J$), dKMC and KMC agree, but as $J$ increases, delocalisation significantly enhances the dKMC efficiency over the classical KMC hopping. For all $J$, overlap CT states separate slightly more efficiently than random CT states and significantly more efficiently than thermalised ones.
    \textbf{b)} IQEs of charge separation starting in random CT states, with $\sigma=\SI{150}{meV}$ and varying $J$, modelled by KMC and dKMC in each dimension. Delocalisation enhances IQE in all dimensions, but more strongly in higher dimensions. The 3D dKMC line stops at a relatively small value of $J$ because of computational cost. The error bars in both panels are the standard errors of the mean.
    }
    \label{fig:results}
\end{figure}

\Cref{fig:results} shows that delocalisation increases the efficiency of charge separation by comparing the IQE calculated with localised polarons, using a standard kinetic Monte Carlo approach (KMC), to that calculated with partially delocalised polarons, using dKMC. When the electronic coupling $J$ is low, the electronic states are localised, meaning KMC and dKMC agree. As $J$ increases, the states become more delocalised and the delocalisation enhancement predicted by dKMC increases regardless of what initial CT states are chosen.

\Cref{fig:results}a shows significantly different efficiencies for thermalised, random, or overlap initial CT states. In both dKMC and KMC, the thermalised CT states separate with a lower efficiency than random CT states, suggesting that, for the parameters studied here, a randomly excited CT state is able to separate before it thermalises. This result supports the idea that hot CT states can improve charge separation in OPVs, as has been observed experimentally~\cite{Bakulin2012,Jailaubekov2013,Grancini2013,Gelinas2014,Falke2014,Tamai2017}. However, our results are also consistent with observations of highly efficient separation from thermalised CT states~\cite{Vandewal2014,Albrecht2014,Kurpiers2018}, because dKMC can reproduce high IQEs from thermalised states, especially if they are assisted by delocalisation.

\begin{figure*}
    \centering
    \includegraphics[width=\textwidth]{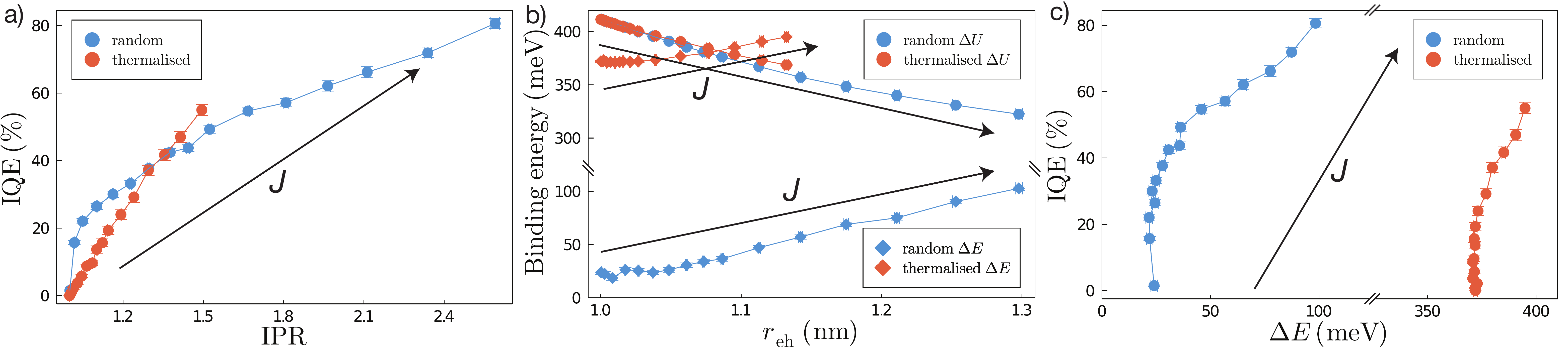}
    \caption{\textbf{Kinetic origin of delocalisation enhancements.}
    \textbf{a)} The charge-separation efficiency (IQE) increases with the delocalisation of the initial CT states ($\mathrm{IPR}$, shown for $J$ from 0 to \SI{75}{meV}). Results are for a 2D heterojunction with $\sigma=\SI{150}{meV}$. 
    \textbf{b)} The efficiency increase with delocalisation is typically attributed to larger initial electron-hole separation $r_\mathrm{eh}$ and the resulting reduction in the Coulomb binding energy $\Delta U$. Although $\Delta U$ does decrease with $r_\mathrm{eh}$, the total binding energy of the initial state ($\Delta E$) actually increases with $r_\mathrm{eh}$ due to delocalisation stabilisation. 
    \textbf{c)} Therefore, the efficiency increases with the total binding energy of the initial state, contrary to the common hypothesis that delocalisation enhancements are simply a consequence of a reduction in binding energy.
    }
    \label{fig:kinetic_effects}
\end{figure*}

These large enhancements can be caused by small amounts of delocalisation. 
\Cref{fig:kinetic_effects}a shows the IQE of random and thermalised CT states as a function of their IPR, illustrating how quickly the efficiency grows with delocalisation. At the highest coupling, $J=\SI{75}{meV}$, thermalised and random CT states have IPRs of only 1.5 and 2.6, but these are enough to
produce five-fold and two-fold IQE enhancements, respectively.
Therefore, just as how small amounts of delocalisation can dramatically improve polaron mobilities~\cite{Balzer2020}, they can also facilitate the separation of charges experiencing a significant Coulomb attraction, a process that is underestimated by standard KMC.

\paragraph{Kinetic effects are more important than the reduction in Coulomb attraction}

Delocalisation has usually been argued to enhance the IQE by increasing the initial electron-hole separation in the CT state, thus reducing the Coulomb attraction and binding energy. 
However, this standard argument is incorrect;
instead, delocalisation enhancements are a kinetic effect chiefly caused by the increased overlap of electronic states (\cref{fig:kinetic_effects}). It is true that the Coulomb binding energy $\Delta U = U_\infty -U_\mathrm{CT} = -U_\mathrm{CT}$ decreases with initial $r_\mathrm{eh}$, as shown in \cref{fig:kinetic_effects}b. However, the Coulomb binding energy is not the sole energetic penalty of charge separation, and \cref{fig:kinetic_effects}b also shows the total binding energy $\Delta E$ as a function of the $r_\mathrm{eh}$ of the initial CT states. It is calculated as $\Delta E = E_\infty -E_\mathrm{CT}$, where $E_\infty$ is the equilibrium energy of two separated polarons, each calculated independently as the thermal expectation value of the energy of a single polaron in a box of size $N_\mathrm{box}^d$.
As shown in \cref{fig:kinetic_effects}b, $\Delta E$ increases with increasing initial delocalisation and $r_\mathrm{eh}$. 

The increase in total binding energy due to delocalisation was predicted by Gluchowski et al.~\cite{Gluchowski2018}, who attributed it to two mechanisms. First, level repulsion between coupled pairs of sites will generally stabilise the lower-energy state, further increasing the binding energy of already most tightly bound states. Second, the lowest-lying CT states generally remain localised as $J$ increases because they are unlikely to have near-resonant neighbours. By contrast, higher-lying CT states become more delocalised at high $J$, increasing their $r_\mathrm{eh}$ until they no longer enter into the averaging for $\Delta U$ and $\Delta E$. Overall, at higher $J$, $\Delta E$ increases because of the greater contribution from low-lying localised traps, which are themselves stabilised by level repulsion.

Overall, \cref{fig:kinetic_effects}c shows that the IQE increases with the binding energy, showing that the delocalisation does not enhance efficiency by reducing the energetic barrier to charge separation.
Instead, delocalisation enhancements must be a purely kinetic (as opposed to energetic) effect, caused by the increased overlaps between the partially delocalised electronic states, which help the polarons move apart faster and further in fewer hops.

\paragraph{Delocalisation enhancements are greater in higher dimensions}

Previously, we found that delocalisation enhances the transport of polarons more in higher dimensions~\cite{Balzer2020}. In particular, polarons are more delocalised in higher dimensions because of the increased number of neighbouring sites, which increase the likelihood of having a neighbouring site with similar energy to allow delocalisation. 

Higher dimensions are also important for accurately modelling charge separation. \Cref{fig:results}b shows that IQEs computed using both regular KMC and dKMC are higher in higher dimensions. In both cases, higher dimensions increase the number of nearest neighbours that polarons can move to, increasing the likelihood of a fast separation pathway. This effect becomes more pronounced when delocalisation is described using dKMC. When the polarons are delocalised, they can move further in one hop, and the number of possible destinations increases with dimension faster than in localised hopping, especially considering the greater delocalisation lengths in higher dimensions~\cite{Balzer2020}.

\begin{figure}
    \centering
    \includegraphics[width=0.95\columnwidth]{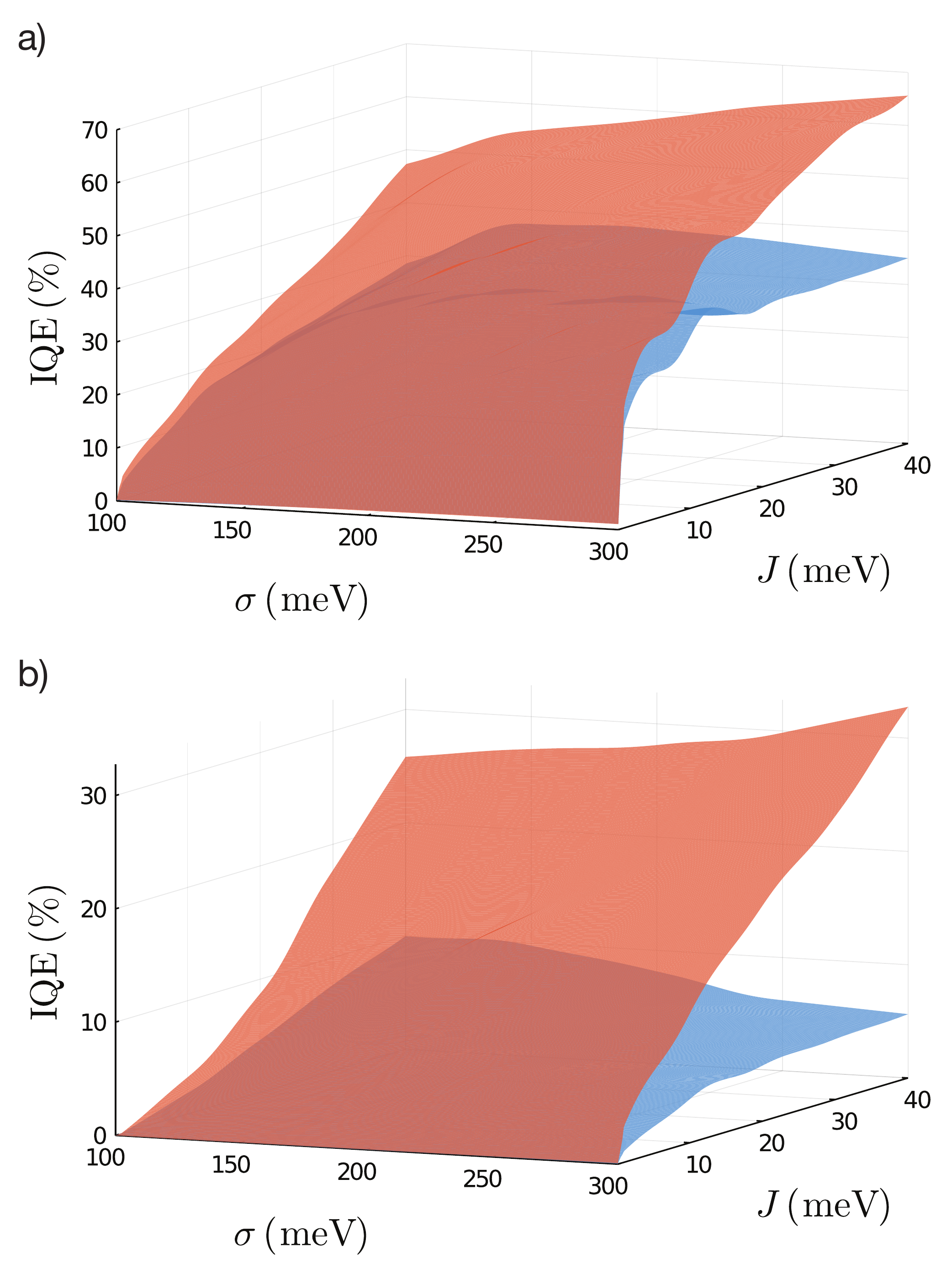}
    \caption{\textbf{Delocalisation enhancements increase with moderate disorder.}
    Parameter scan of internal quantum efficiencies (IQEs) of charge separation in 2D for \textbf{a)} random and \textbf{b)}  thermalised CT states, as a function of electronic coupling $J$ and disorder $\sigma$, using KMC (blue) and dKMC (orange). dKMC agrees with KMC in the low-$J$ limit (when the states are localised), and always predicts IQEs greater than KMC as $J$ increases. For both random and thermalised CT states, higher $J$ always leads to higher IQE; the behaviour as a function of disorder is more complicated, although modest amounts of disorder generally increase the IQE.}
    \label{fig:scan}
\end{figure}

\paragraph{Delocalisation enhancements increase with moderate disorder}

dKMC's computational savings allow parameter scans that independently probe the roles of microscopic parameters. \Cref{fig:scan} shows the IQE as a function of both the electronic coupling $J$, which increases delocalisation, and energetic disorder $\sigma$, which localises states. In the limit of small $J$, when the electronic states are localised onto individual sites, dKMC and KMC agree. As $J$ increases, causing delocalisation, the efficiency of separation is always greater when delocalisation is included using dKMC.

\Cref{fig:scan} reveals the subtle effect of disorder on IQE. It is known that a modest amount of disorder can be beneficial for the separation of localised charges~\cite{Shi2017,Jankovic2018,Athanasopoulos2019} because it provides energetically downhill pathways for the separation of many initial CT states.
\Cref{fig:scan} shows that delocalisation extends the regime over which disorder is beneficial for charge separation, because it increases the overlap between electronic states allowing for polarons to move further and faster. 
For extreme disorders, the delocalisation enhancement eventually decreases as the states become completely localised, giving agreement between dKMC and KMC.

\section{Discussion}
\paragraph{Computational limitations}

Despite the savings enabled by dKMC, it remains limited by computational cost. Especially in three dimensions, dKMC is limited to smaller values of the electronic coupling (\cref{fig:results}b) because the polaron states become more delocalised at higher $J$, requiring the diagonalisation of large Hamiltonians. This problem is particularly acute in the two-particle Hilbert space, where Hamiltonian diagonalisation scales as $O(N_\mathrm{box}^{18})$ in three dimensions. However, the results in \cref{fig:results}b already reveal significant enhancements at low $J$, especially compared to lower dimensions, and we expect that the enhancement would continue increasing at higher couplings. Future computational advances may extend the results to higher $J$.

\paragraph{Possible extensions}

In the future, dKMC could be extended to address other important questions in the physics of disordered materials. We envisage both extensions that relax some of our assumptions as well as applications to novel physical problems.

Most assumptions in the current version of dKMC could be relaxed. Doing so would likely lead to an increase in the predicted role of delocalisation, because all the assumptions in this work are conservative, chosen to underestimate the extent and effects of delocalisation.

The simplest extension would be the inclusion of non-nearest-neighbour couplings $J_{mn,m'n}$ and $J_{mn,mn'}$ in \cref{eq:H_S}. Doing so would not require any alterations to the dKMC algorithm, but would result in more delocalised polaron states, assuming the nearest-neighbour couplings are kept constant.

More generally, while we studied parameter ranges where the underlying sPTRE master equation is valid, more general master equations could be used in other cases. 
In particular, sPTRE can be inaccurate for systems weakly coupled to slow baths, when the secular and Markov approximations fail~\cite{Lee2012,Chang2013,Pollock2013,Lee2015}.
In those cases, electronic dynamics can be faster than bath relaxation, meaning that the bath modes are not fully relaxed, as assumed in \cref{eq:polaron}.
dKMC could be applied to those systems by replacing the fully displaced polaron transformation with its variational counterpart~\cite{Silbey1984,Zimanyi2012,Pollock2013,Jang2022}, which would also allow Ohmic or sub-Ohmic baths to be treated. 

Similarly, the secular approximation inherent in sPTRE could be relaxed. The secular approximation leads to a neglect of coherences between polaronic states, which reduces the number of elements of the density matrix being tracked from $O(N^{2d})$ to $O(N^{d})$. In the regime we considered, this is a good approximation~\cite{Lee2015}, because coherences between charge states are rarely induced and almost never survive long enough to affect long-time charge transport. Nevertheless, dKMC could be expanded to include coherences if desired and if the computational resources allow it. 

A final pair of assumptions in our treatment is that both the static disorder (\cref{eq:H_S}) and the system-bath coupling (\cref{eq:H_SB}) are diagonal in the site basis. More generally, the Hamiltonian could be modified to include both off-diagonal disorder and off-diagonal system-bath couplings. The latter are particularly important in organic crystals, where the fluctuations of electronic couplings are an additional source of localisation of the electronic states~\cite{Troisi2006, Troisi2006_Review}. Of the two changes, off-diagonal disorder could be implemented without any changes to the dKMC algorithm by replacing the constant $J$ with a site-dependent one. By contrast, off-diagonal system-bath coupling would require modifications to the equations of motion. Polaron theories generally rely on the ability of the polaron transformation to eliminate diagonal system-bath couplings; therefore, they are unable to fully eliminate off-diagonal ones, and the remainder would have to be included in the perturbative treatment.

Two-body dKMC could also be extended beyond the separation of CT states. In particular, adding  excitonic states to the simulation would give a quantum treatment of exciton dissociation both in the bulk and at interfaces.
It could also describe exciton-to-CT-state transfer and predict the nature of CT states that mediate charge separation, obviating the need for the comparison between the random, thermalised, and overlap CT states introduced above. Doing so may explain and unite the seemingly disparate experimental observations of efficient separation requiring hot CT states and those occurring from thermalised CT states. The inclusion of excitons would, however, pose computational challenges, because it would require replacing nearest-neighbour electronic couplings with long-range excitonic ones. 

A process similar to exciton dissociation is singlet fission, and we expect that an excitonic dKMC code could be adapted to treat that problem as well, replacing the separated polarons with separated triplets.

Finally, we anticipate that it will be possible to use dKMC to parametrise drift-diffusion simulations by computing the necessary dissociation and recombination rates. Doing so would allow a multiscale simulation of a broad range of processes in organic semiconductors in a way that takes quantum effects into account.

\paragraph{Conclusions}
Our results explain how even small amounts of delocalisation substantially assist polarons in escaping their large Coulomb attraction in OPVs. This is largely a kinetic effect, caused by increased overlap of electronic states, as opposed to the common hypothesis of a reduction in Coulomb binding. For instance, initial delocalisation across less than two molecules can suffice to make the separation of thermalised CT states five times more efficient. Furthermore, higher-dimensional effects, including greater delocalisation, are particularly important to capturing the physics of charge separation. 
These results have been made possible by dKMC, a robust and computationally efficient technique for modelling charge-separation dynamics that includes all of the important features: three dimensions, delocalisation, disorder, and polaron formation. Overall, dKMC opens new avenues in the exploration of organic semiconductors by allowing a more comprehensive theoretical description than has been possible.

\section{\label{sec:theory}Methods}

\begin{figure*}
    \centering
    \includegraphics[width=0.9\textwidth]{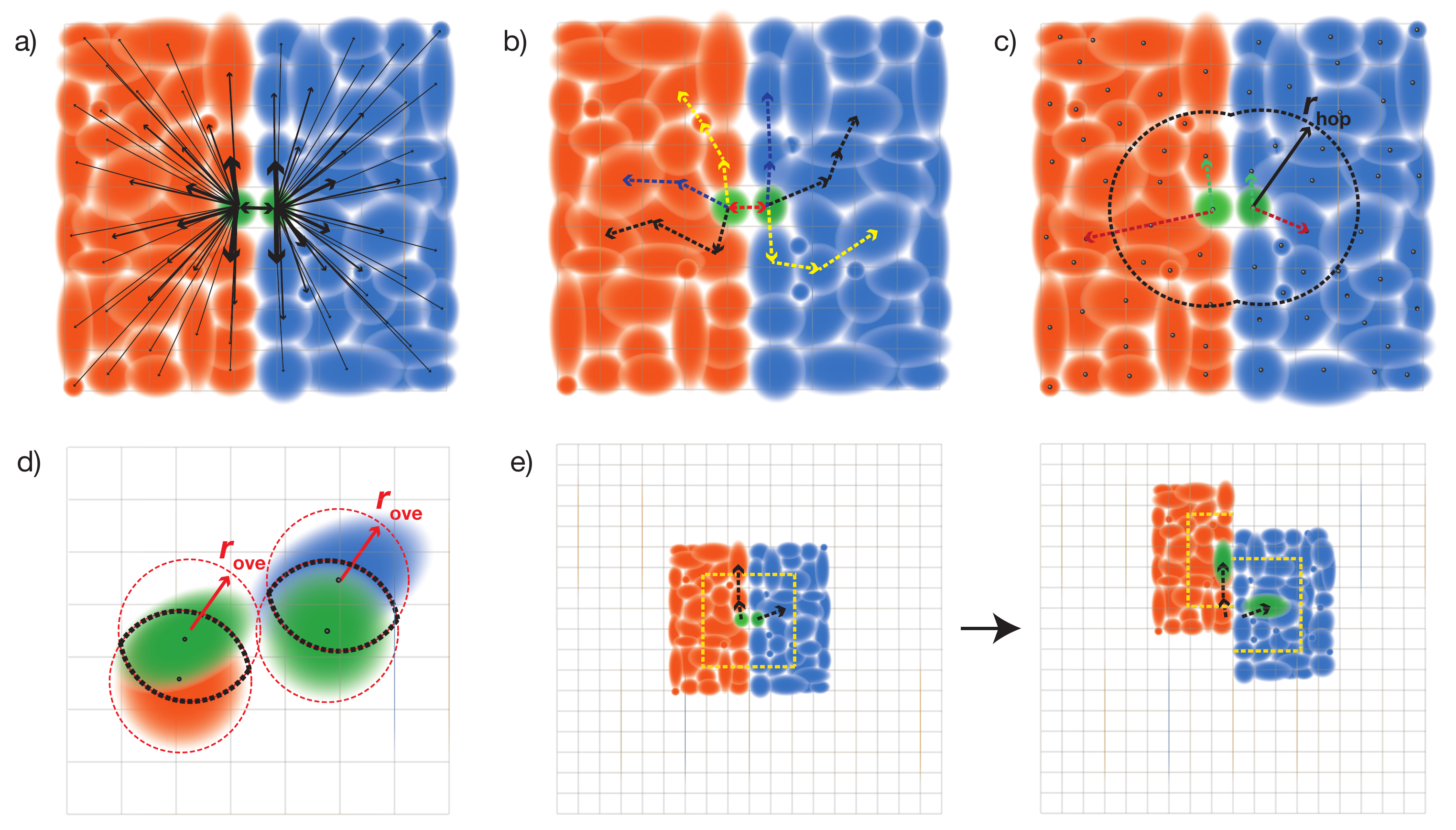}
    \caption{
    \textbf{Computational components of dKMC for charge separation.}
    \textbf{a)} The full sPTRE master equation tracks the time-dependent evolution of the full charge density of the electron and hole through all polaron states. To avoid the computational cost of doing so, we apply four approximations (panels b--e).
    \textbf{b)} Kinetic Monte Carlo: we average many trajectories, formed probabilistically from sequential hops.
    \textbf{c)} Hopping radius $r_\mathrm{hop}$: we only calculate rates to polaron states where the centre (black dots) of the electron and hole are close enough.
    \textbf{d)} Overlap radius $r_\mathrm{ove}$: when calculating transfer rates between polaron states, we only consider pairs of sites (lattice points) that are close to both states. 
    \textbf{e)} Diagonalising on the fly: rather than diagonalising the full Hamiltonian, we diagonalise a subsystem surrounding each polaron. Once either polaron moves too close to the edge of its box, we diagonalise a new subsystem.}
    \label{fig:approximations}
\end{figure*}

Here, we extend dKMC from the one-particle mobility calculation~\cite{Balzer2020} to the two-particle charge-separation problem by applying the polaron transformation to the Hamiltonian, establishing the sPTRE master equation for the two-particle picture, and then establishing the dKMC procedure for charge separation.

\paragraph{Polaron transformation} 
Applying the polaron transformation to the total Hamiltonian (\cref{eq:H}) yields
\begin{equation}
\tilde{H} = e^S He^{-S} = \tilde{H}_\mathrm{S} + \tilde{H}_\mathrm{B} + \tilde{H}_\mathrm{SB}.
\end{equation}
The polaron-transformed system Hamiltonian is
\begin{multline}
    \label{eq:polaron_H_S}
    \tilde{H}_\mathrm{S} = \sum_{m\in D,n\in A} \tilde{E}_{mn} \ket{m,n}\bra{m,n} \\ + \sum_{m \neq m' \in D,n\in A} \kappa_{mm'}J_{mn,m'n} \ket{m,n}\bra{m',n} \\ + \sum_{m\in D,n \neq n'\in A} \kappa_{nn'}J_{mn,mn'} \ket{m,n}\bra{m,n'}
\end{multline}
where $\tilde{E}_{mn}=E_{mn}-\sum_k\abs{g_{mk}}^2/\omega_{mk}-\sum_k\abs{g_{nk}}^2/\omega_{nk}$, and the electronic couplings are renormalised by
\begin{multline}
    \label{eq:large_kappa}
    \kappa_{mm'} = e^{-\frac{1}{2}\sum_k\left(\frac{g^2_{mk}}{\omega^2_{mk}}\coth{\frac{\beta\omega_{mk}}{2}}+\frac{g^2_{m'k}}{\omega^2_{m'k}}\coth{\frac{\beta\omega_{m'k}}{2}}\right)},
\end{multline}
where $\beta=1/k_\mathrm{B} T$ and we take $T=\SI{300}{K}$.
The bath Hamiltonian is unchanged, $\tilde{H}_\mathrm{B}=H_\mathrm{B}$, while the polaron-transformed interaction Hamiltonian is
\begin{multline}
    \label{eq:polaron_H_SB}
    \tilde{H}_\mathrm {SB} = \sum_{m \neq m' \in D,n\in A} J_{mn,m'n} \ket{m,n}\bra{m',n}V_{mm'} \\ + \sum_{m\in D,n \neq n'\in A}J_{mn,mn'} \ket{m,n}\bra{m,n'}V_{nn'},
\end{multline}
where
\begin{multline}
    V_{mm'} = \exp{\left(\sum_k\frac{g_{mk}}{\omega_{mk}}\left(b^\dag_{mk}-b_{mk}\right)\right)}\\ \exp{\left(-\sum_k\frac{g_{m'k}}{\omega_{m'k}}\left(b^\dag_{m'k}-b_{m'k}\right)\right)}- \kappa_{mm'}.
\end{multline}

We apply two standard approximations to reduce the computational cost of summing over many bath modes: all sites couple to their baths with the same strength, $g_{nk}=g_k$, and the spectral density $J(\omega)=\sum_k g_k^2\delta(\omega-\omega_k)$ is replaced with a continuous function. In principle, one could choose structured spectral densities for specific organic molecules; here, we choose the widely used super-Ohmic spectral density $J(\omega) = \frac{\lambda}{2} (\omega/\omega_c)^3 \exp(-\omega/\omega_c)$~\cite{Jang2002,Jang2011,Pollock2013,Wilner2015, Lee2015}, where $\lambda=\SI{100}{meV}$ is the reorganisation energy and $\omega_{c}=\SI{62}{meV}$~\cite{Lee2015} is the cutoff frequency. 

The two approximations simplify the renormalisation factors in \cref{eq:large_kappa} to
\begin{equation}
    \kappa_{mm'} = \kappa = e^{-\int_0^\infty d\omega\frac{J(\omega)}{\omega^2}\coth{\left(\beta\omega/2\right)}}.
\end{equation}
Because $\kappa<1$, it always reduces the electronic coupling in \cref{eq:polaron_H_S}, meaning that the polaron transformation reduces the delocalisation of the electronic states~\cite{Rice2018}.

\paragraph{Secular Redfield theory} 

We use the secular polaron-transformed Redfield master equation (sPTRE)~\cite{Lee2015}. As the polaron transformation reduces the system-bath coupling, we apply the second-order perturbative Redfield theory to $\tilde{H}_\mathrm{SB}$~\cite{Lee2015}. Following the secular approximation, which is accurate for most disordered materials of interest~\cite{Lee2015, Balzer2020}, we obtain the sPTRE master equation, 
\begin{equation}
    \label{eq:sPTRE_master_equation}
    \frac{d\rho_{\nu}(t)}{dt} = \sum_{\nu'}R_{\nu\nu'}\rho_{\nu'}(t),
\end{equation}
which describes the evolution of the populations of polaron states, found by diagonalising $\tilde{H}_\mathrm{S}$. This evolution is determined by the secular Redfield tensor $R_{\nu\nu'}$,
\begin{equation}
    \label{eq:secular_redfield_tensor}
    R_{\nu\nu'} = 2\Re\Big(\Gamma_{\nu'\nu,\nu\nu'} - \delta_{\nu\nu'}\sum_\kappa\Gamma_{\nu\kappa,\kappa\nu'}\Big),
\end{equation}
where the damping rates are
\begin{multline}
 \label{eq:Gamma}
     \Gamma_{\mu\nu,\mu'\nu'} = \sum_{m \in D}\sum_{p,q,p',q' \in A} J_{mp,mq} J_{mp',mq'} \braket{\mu|m,p} \\ \braket{m,q|\nu}\braket{\mu'|m,p'}\braket{m,q'|\nu'}K_{pq,p'q'}(\omega_{\nu'\mu'}) \\ 
     + \sum_{n \in A}\sum_{p,q,p',q' \in D} J_{pn,qn} J_{p'n,q'n} 
    \braket{\mu|p,n}
    \\\braket{q,n|\nu}\braket{\mu'|p',n}\braket{q',n|\nu'} K_{pq,p'q'}(\omega_{\nu'\mu'}),
\end{multline}
with $\omega_{\nu'\mu'}=E_{\nu'}-E_{\mu'}$ and
\begin{equation}
\label{eq:int_bath_cor}
K_{pq,p'q'}(\omega) = \int_0^\infty e^{i\omega \tau}\braket{\tilde{\hat{V}}_{pq}(\tau)\tilde{\hat{V}}_{p'q'}(0)}_B d\tau.
\end{equation}
The bath correlation function, where the hats denote the interaction picture, is given by~\cite{Jang2011}
\begin{equation}
\braket{\tilde{\hat{V}}_{pq}(\tau)\tilde{\hat{V}}_{p'q'}(0)}_B = \kappa^2(e^{\lambda_{pq,p'q'}\phi(\tau)} - 1),
\end{equation}
where $\lambda_{pq,p'q'} = \delta_{pp'} - \delta_{pq'} + \delta_{qq'} - \delta_{qp'}$
and 
\begin{equation}
\phi(\tau) = \int_0^\infty d\omega\frac{J(\omega)}{\omega^2}\big(\cos(\omega \tau)\coth(\beta\omega/2) - i \sin(\omega \tau)\big).
\end{equation}

\begin{algorithm}
    \fbox{
    \begin{minipage}{0.45\textwidth}
        \setlist{nolistsep}
        \raggedright
        Given parameters $N$, $d$, $\sigma$, $J$, $\lambda$, $\omega_c$, $T$, $R_\mathrm{recomb}$, $r_\mathrm{sep}$, $n_\mathrm{hops}$, $n_\mathrm{iter}$, and $n_\mathrm{traj}$:
        \begin{enumerate}[leftmargin=*]
            \item Calculate $r_\mathrm{hop}$ and $r_\mathrm{ove}$ as described in \cite{Balzer2020}.
            \item For $n_\mathrm{iter}$ realisations of disorder:
            \begin{enumerate}[leftmargin=*, label=\alph*.]
                \item Generate $N^d$ lattice of sites, with $N^d/2$ random donor HOMO energies and $N^d/2$ random acceptor LUMO energies.
                \item Create a polaron-transformed $\tilde{H}_S$ containing pairs of sites within a box of size $N_\mathrm{box}^d=(r_\mathrm{ove}+r_\mathrm{hop})^d$ at the centre of the lattice. Diagonalise $\tilde{H}_S$ to find the polaron states, their energies, and the positions of electrons and holes in every state.
                \item For $n_\mathrm{traj}$ trajectories:
                \begin{enumerate}[leftmargin=*]
                    \item Set $n_\mathrm{sep}\leftarrow 0$ and choose an initial state $\nu$ in any of the ways described in the text.
                    \item For $n_\mathrm{hops}$ hops:
                    \begin{enumerate}[leftmargin=*]
                        \item Create a list $L$ of all states $\nu'$ such that  $\abs{\vec{C}^e_\nu-\vec{C}^e_{\nu'}} + \abs{\vec{C}^h_\nu-\vec{C}^h_{\nu'}} < r_\mathrm{hop}$.
                        \item Calculate $R_{\nu\nu'}$ for all $\nu'\in L$ using \cref{eq:secular_redfield_tensor}, neglecting all terms in  \cref{eq:Gamma} that contain overlaps of the form $\braket{\alpha|x,y}$ such that $\abs{\vec{r}_x-\vec{C}^e_{\alpha}} + \abs{\vec{r}_y-\vec{C}^h_{\alpha}} > r_\mathrm{ove}$.
                        \item Calculate $k^\nu_\mathrm{recomb}$ using \cref{eq:recomb} and append $g$ to $L$.
                        \item Set $S_{\nu'}\leftarrow\sum_{\mu=1}^{\nu'} R_{\nu\mu}$ for all $\nu'\in L$ and set $T\leftarrow \sum_{\nu'\in L}S_{\nu'}$.
                        \item Find $\nu'$ such that $S_{\nu'-1} < uT < S_{\nu'}$, for uniform random number $u \in (0,1]$, and update $\nu\leftarrow\nu'$. 
                        \item If $\nu=g$, exit the for loop.
                        \item If $\abs{\vec{C}^e_\nu-\vec{C}^h_\nu}>r_\mathrm{sep}$, set $n_\mathrm{sep}\leftarrow n_\mathrm{sep}+1$, and exit the for loop.
                        \item If $\vec{C}^e_\nu$ or $\vec{C}^h_\nu$ is within $N_\mathrm{box}/2$ of the edge of the current boxes, diagonalise a new $\tilde{H}_S$ containing pairs of sites within two boxes of size $N_\mathrm{box}^d$ centred at $\vec{C}^e_\nu$ and $\vec{C}^h_\nu$.
                    \end{enumerate}
                \end{enumerate}
                 \item Calculate $\mathrm{IQE}=n_\mathrm{sep}/n_\mathrm{traj}$.
            \end{enumerate}
            \item Calculate mean IQE by averaging all IQEs.
        \end{enumerate}
    \end{minipage}
    }
    \caption{dKMC for charge separation.}
    \label{alg:listing}
\end{algorithm}

\paragraph{dKMC} In principle, the full evolution of the density matrix of both the electron and hole polarons could be tracked using the sPTRE master equation. However, doing so is expensive for three reasons. First, finding the polaron states requires diagonalising $\tilde{H}_S$, which scales as $O(N^{6d})$. Second, calculating all components of the Redfield tensor (\cref{eq:secular_redfield_tensor}) requires rates between all pairs of states, of which there are  $O(N^{4d})$. Finally, calculating each of these rates requires computing the damping rates (\cref{eq:Gamma}), each of which includes a sum over $O(N^{5d})$ terms. Therefore, propagating the full sPTRE master equation would scale roughly as $O(N^{6d})+O(N^{9d})$.

To reduce the complexity, we use the four approximations developed for the original dKMC method~\cite{Balzer2020}. Summarised in \cref{fig:approximations} and \cref{alg:listing}, they are: 
\begin{enumerate}[topsep=0pt,itemsep=0pt,parsep=0pt,partopsep=0pt]
    \item dKMC maps the master equation onto kinetic Monte Carlo (\cref{fig:approximations}b). Instead of propagating the full density matrix, it tracks, and averages over, $n_\mathrm{traj}$ stochastic trajectories. The trajectories are found by hopping from one state to another, with the target chosen probabilistically in proportion to the corresponding Redfield rate. This procedure is continued until the polarons recombine, become separated by a distance $r_\mathrm{sep}$, or hop more than $n_\mathrm{hops}$ times. The trajectory approach means only outgoing rates need to be calculated before each hop, rather than between all pairs of states, reducing the number of rate to be calculated from $O(N^{4d})$ to $O(N^{2d}n_\mathrm{hops}n_\mathrm{traj})$. The IQE is the percentage of the trajectories where polarons separate, averaged over $n_\mathrm{iters}$ realisations of disorder. 

    \item Instead of calculating all outgoing rates from the current state, dKMC only calculates rates to states where the combined distance that the electron and holes hop is less than a hopping radius $r_\mathrm{hop}$ (\cref{fig:approximations}c). This approximation is justified by the small spatial overlap, and therefore Redfield rate, between states where either the electron or the hole (or both) hop far away. To calculate the combined hopping distance, we consider the positions of the electron and hole in state $\nu$ to be the expectation values $\vec{C}^e_\nu=\bra{\nu}\vec{r}^e\ket{\nu}$ and $\vec{C}^h_\nu=\bra{\nu}\vec{r}^h\ket{\nu}$, and only calculate rates from state $\nu$ to states $\nu'$ if $\abs{\vec{C}^e_\nu-\vec{C}^e_{\nu'}} + \abs{\vec{C}^h_\nu-\vec{C}^h_{\nu'}} < r_\mathrm{hop}$. Doing so reduces the number of rates calculated at each hop from $O(N^{2d})$ to $O(r_\mathrm{hop}^{2d})$. We use $r_\mathrm{hop}$ values benchmarked for single-particle dynamics~\cite{Balzer2020}. The approximation is controlled by increasing or decreasing $r_\mathrm{hop}$ to achieve the target accuracy.

    \item When calculating each rate, instead of summing over $N^{5d}$ quintuples ($m\in D$ and $p,q,p',q'\in A$, or $n\in A$ and $p,q,p',q'\in D$) in \cref{eq:Gamma}, we only sum over a truncated set (\cref{fig:approximations}d), because Anderson localisation of the wavefunctions makes overlaps between distant states small. 
    To do so, we neglect all terms in \cref{eq:Gamma} that contain overlaps of the form $\braket{\alpha|x,y}$ such that $\abs{\vec{r}_x-\vec{C}^e_{\alpha}} + \abs{\vec{r}_y-\vec{C}^h_{\alpha}} > r_\mathrm{ove}$, where $\alpha$ is one of $\mu,\nu,
    \mu'$, or $\nu'$ and $x$ and $y$ are any of $m,n,p,q,p'$, or $q'$.
    This reduces the number of terms in each rate calculation from $~O(N^{5d})$ to $~O(r_\mathrm{ove}^{5d})$. The accuracy is controllable by adjusting $r_\mathrm{ove}$; we use $r_\mathrm{ove}$ values calculated in \cite{Balzer2020}.

    \item Instead of diagonalising the entire Hamiltonian to generate polaron states, we diagonalise on the fly smaller Hamiltonians encompassing pairs of sites within boxes of size $N_\mathrm{box}^d$ surrounding both polarons (\cref{fig:approximations}e). The polarons can move within their boxes until either one gets too close to the edge of its box, when the boxes are moved and rediagonalised. As the computational bottleneck is often diagonalising large Hamiltonians, we make the boxes as small as possible, choosing $N_\mathrm{box}=(r_\mathrm{hop}+r_\mathrm{ove})$, and rediagonalise when either polaron is within $N_\mathrm{box}/2$ of the edge of its box. This reduces the diagonalisation cost from $O(N^{6d})$ to $O(N_\mathrm{box}^{6d}n_\mathrm{rediag})$, for $n_\mathrm{rediag}$ rediagonalisations. 
\end{enumerate}

Overall, dKMC reduces the computational complexity from $O(N^{6d})+O(N^{9d})$ to $O(N_\mathrm{box}^{6d}n_\mathrm{rediag}) + O(r_\mathrm{hop}^{4d}r_\mathrm{ove}^{5d}n_\mathrm{hops}n_\mathrm{traj})$.

\paragraph{Recombination} In standard kinetic Monte Carlo, CT-state recombination is often described with a constant rate $R_\mathrm{recomb}$, occurring when the electron and hole are on adjacent sites across the donor-acceptor interface. For dKMC, we calculate the recombination rate of CT state $\nu$ using Fermi's golden rule,
\begin{equation}
    k^\nu_\mathrm{recomb}=2\pi\Bigg|\sum_{(m,n)\in \mathrm{CT}}\braket{\nu|m,n}\braket{m,n|H|g}\Bigg|^2\rho_\mathrm{recomb},
\end{equation}
where the sum goes over interfacial site pairs and $\rho_\mathrm{recomb}$ is the density of states. Assuming the interfacial pairs are coupled to the ground state with constant coupling $\braket{m,n|H|g}=J_\mathrm{recomb}$, the rate becomes
\begin{equation}
    \label{eq:recomb}
    k^\nu_\mathrm{recomb}= R_\mathrm{recomb}\Bigg|\sum_{(m,n)\in \mathrm{CT}}\braket{\nu|m,n}\Bigg|^2,
\end{equation}
where $R_\mathrm{recomb}=2\pi|J_\mathrm{recomb}|^2\rho_\mathrm{recomb}$.
Recombination therefore occurs at the ordinary Monte-Carlo rate modified by the square of the sum of the CT state's amplitudes on all interfacial site pairs. This result agrees with generalised Marcus theory for transfer between weakly coupled delocalised states~\cite{Taylor2018}, and with previous recombination studies~\cite{Tempelaar2016}.
In calculations, we used $R_\mathrm{recomb}=\SI{e-10}{s^{-1}}$.

\begin{acknowledgments}
We were supported by the Westpac Scholars Trust (Research Fellowship and Future Leaders Scholarship), by the Australian Research Council (DP220103584), by the Australian Government Research Training Program, and by the University of Sydney Nano Institute Grand Challenge \textit{Computational Materials Discovery}.
We were supported by computational resources from the National Computational Infrastructure (Gadi) and by the University of Sydney Informatics Hub (Artemis).
\end{acknowledgments}

\end{document}